# Dispersion Splitting of Phonon Polaritons in van der Waals Heterostructure

*Daeho Noh[†], Jaehyeong Ock[†], Sergey G. Menabde\*, Min Seok Jang\*\**

School of Electrical Engineering, Korea Advanced Institute of Science and Technology, Daejeon 34141, Korea

\*menabde@kaist.ac.kr

\*\*jang.minseok@kaist.ac.kr

[†]These authors contributed equally to this work.

ABSTRACT: The biaxial van der Waals crystal α-phase molybdenum trioxide (α-$MoO_3$) supports hyperbolic phonon-polaritons with anomalous dispersion in the Type-I Reststrahlen band (RB-I). Despite the low loss and long lifetime of these polaritons, dispersion engineering in this regime has remained largely unexplored. In this work, we show that when two α-$MoO_3$ slabs are placed in close proximity, their eigenmodes hybridize and the dispersion splits into two branches with different momenta and field symmetry, providing a powerful platform for dispersion manipulation. We experimentally demonstrate the polaritonic mode splitting in α-$MoO_3$ within a heterostructure with hexagonal boron nitride (hBN) employed as a spacer, probed by a scattering-type scanning near-field optical microscope. Furthermore, we propose a design framework for active and mode-selective tailoring of the polaritonic dispersion in the heterostructure incorporating graphene, achieved through tuning its Fermi energy. Our work experimentally demonstrates the feasibility of phonon-polariton mode splitting in the RB-I and suggests a new platform for dispersion engineering of hyperbolic phonon-polaritons in general.

**Introduction**

Polaritons are quasiparticles that arise from the coupling of light to charge oscillations in matter. They typically possess much larger momentum than free-space photons, exhibit high field confinement, and thus can overcome the diffraction limit of conventional optical systems. Among them, hyperbolic phonon-polaritons (PhPs), which emerge from the coupling between the atomic lattice vibrations (phonons) and photons, have been observed in many van der Waals (vdW) materials[1] such as alpha-phase molybdenum trioxide[2] ($\alpha$-MoO$_3$) and hexagonal boron nitride (hBN)[3]. Depending on the transverse and longitudinal phonon energies, the sign of at least one of the components of the material's permittivity tensor may become negative within the specific frequency range called the Reststrahlen band (RB). Generally, there are two types of RB: Type I RB (RB-I) where the out-of-plane permittivity component is negative, and Type II RB (RB-II) where at least one in-plane permittivity component is negative. Phonon-polaritons in the RB-I exhibit anomalous dispersion and negative group velocity, while polaritons in the RB-II display positive group velocity and normal (conventional) dispersion.[4]

Notably, $\alpha$-MoO$_3$ is a biaxial crystal, so PhPs exhibit in-plane anisotropy. In addition to $\alpha$-MoO$_3$, various other two-dimensional vdW materials with in-plane anisotropy, such as $\beta$-Ga$_2$O$_3$, have been discovered[5]. Stacked and twisted heterostructures of these materials have demonstrated exotic properties, prompting active research efforts aimed at engineering the polaritonic dispersion.[6] At the same time, when hBN and $\alpha$-MoO$_3$ are stacked on a gold crystal, they support high-momentum image modes.[7, 8] Furthermore, by introducing a spacer between the gold and hBN, the manifestation of first-order symmetric and second-order antisymmetric modes has been demonstrated.[9] Recently, a polaritonic Fourier crystal with a spatially varying gap between hBN (or $\alpha$-MoO$_3$) and gold has been shown to induce a bandgap in the PhP dispersion.[10, 11] In graphene/$\alpha$-MoO$_3$ or graphene/hBN heterostructures, the PhP dispersion can be modulated by tuning the Fermi energy of graphene.[12,13] Other approaches to control polariton dispersion include angle-dependent dispersion[14] in $\beta$-Ga$_2$O$_3$, and chemical intercalation that modulates the intrinsic properties of vdW crystals.[15]

However, no study has experimentally demonstrated PhP hybridization and splitting into bonding and antibonding modes in a system of two (not twisted) layers of the same polaritonic material separated by a narrow gap or a spacer. Using two identical vdW slabs instead of the metallic mirror substrate enables the existence of additional modes with opposite field symmetry that are not allowed by mirror charges in the metal that produce image modes. As the gap between the slabs decreases, coupling between their eigenmodes arises, causing the splitting of two degenerate modes into symmetric and an anti-symmetric modes (Figure 1A), where symmetry refers to the out-of-plane component of the electric field. We refer to this phenomenon as "mode splitting". Importantly, such splitting results in modes with converging momenta at a narrowing gap, in contrast to image modes.[9] When the gap vanishes entirely, the system converges to a single slab with twice the thickness, and the symmetric and antisymmetric modes converge to those of orders $N$ and $N+1$ (Figure 1A).

Here, we report the experimental observation of the splitting of PhP modes in an α-MoO$_3$/hBN/α-MoO$_3$ heterostructure, with a focus on the RB-I of α-MoO$_3$. We extract the PhP dispersion from the near-field interference patterns using the scattering-type scanning near-field optical microscope (s-SNOM), and compare the experimental results with numerical solutions based on the transfer matrix method (TMM).[16] Furthermore, we numerically demonstrate various ways to manipulate the PhP dispersion in the heterostructure by varying the spacer and slab thickness, including structural symmetry. Finally, we propose a design framework for active tailoring of the polaritonic dispersion in the heterostructure with graphene where controlled mode splitting is achieved through tuning the graphene Fermi energy.

**Results**

When two identical slab waveguides are infinitely separated, their eigenmodes are degenerate and independent of each other. However, as the distance between the slabs is reduced, the presence of the second slab can no longer be neglected. The permittivity of each slab then acts as a perturbation to the eigenvalue problem of the other slab. Akin to the splitting of degenerate eigenstates of two neighboring hydrogen atoms into bonding and anti-bonding states, the eigenmodes of the two adjacent waveguides in electrodynamics are expected to split into two non-degenerate modes with opposite electric field symmetry. This phenomenon can be understood as a result of mode interaction through evanescent fields beyond the slabs.[17]

In this work, we use the mid-infrared PhP in α-MoO$_3$ to study the mode splitting phenomenon. First, we consider an ideal situation where the two slabs are suspended in air and separated by an air gap. In Figure 1B, the PhP dispersion in a suspended α-MoO$_3$ slab is shown for the RB-I and the upper RB-II frequencies, and Figure 1C shows the PhP dispersion in the two adjacent slabs separated by 40 nm of air. In the latter case, each PhP mode splits into symmetric and antisymmetric modes, regardless of the RB type and mode order. Figure S1A in the Supplementary Information shows the dispersion splitting evolution in RB-I with respect to the gap size. The symmetry of the split modes can be easily tracked as the symmetric mode always has a lower momentum in RB-I. However, there are three major differences between the dispersion behavior in different RBs.

First, the electric field symmetry between the split modes is different. As shown in Supplementary Figure S2A, when the out-of-plane electric field is considered, the fundamental mode of a single α-MoO$_3$ slab in RB-I is a symmetric mode. For the two adjacent slabs, this mode splits into a lower-momentum symmetric and higher-mometum anti-symmetric modes. In contrast, in RB-II, the fundamental mode of a single slab is an anti-symmetric mode, and in the two adjacent slabs, it splits into a lower-momentum anti-symmetric and a higher-momentum symmetric mode.

Second, when considered at the same frequency, the distance between the split modes in momentum space depends on the gap size. Interestingly, the mode separation in RB-I is much more sensitive to the gap compared to RB-II. This indicates that more efficient dispersion manipulation is possible in RB-I rather than in RB-II (Figure 1D, top panel).

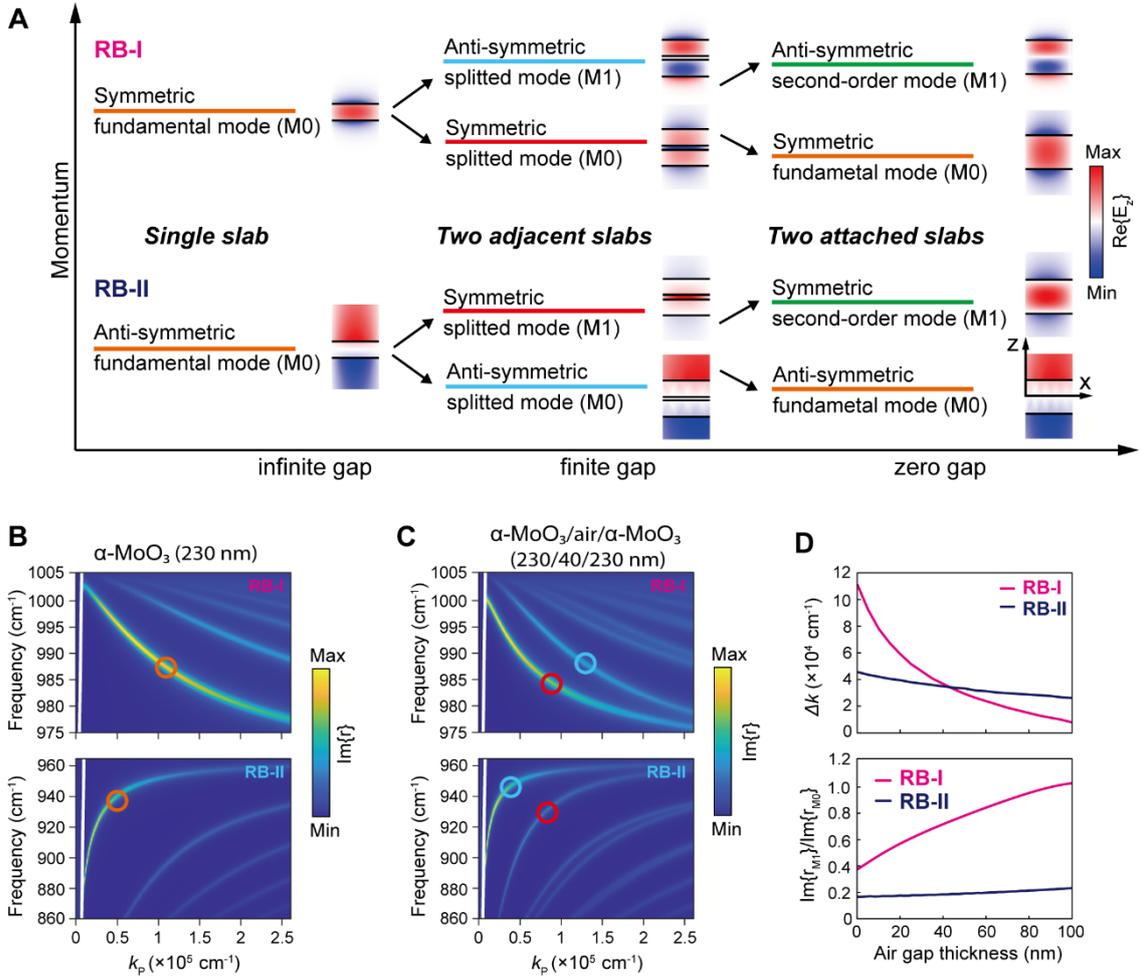

**Figure 1. (A)** Schematics of the momentum and field distribution of the eigenmodes in the two adjacent α-MoO$_3$ slabs as a function of their separation distance in RB-I (top row) and RB-II (bottom row). The out-of-plane component of the electric field $E_z$ illustrates the mode symmetry. **(B)** Dispersion of PhP modes in a single 230 nm-thick free-standing α-MoO$_3$ slab. **(C)** PhP dispersion in the two adjacent slabs of the same thickness separated by a 40 nm air gap. PhP dispersions in (B) and (C) are calculated using the TMM; the white line represents the light line, and the circles indicate corresponding modes in (A). **(D)** Top panel: momentum difference between the symmetric and anti-symmetric split modes in the two-slab system as a function of the air gap thickness. Bottom panel: ratio of the maximal imaginary part of the reflection coefficient calculated for the M0 and M1 modes as a function of the air gap thickness. The data is calculated for the central frequencies of RB-I and RB-II: 985 cm$^{-1}$ (pink) and 893 cm$^{-1}$ (purple), respectively.

Finally, the relative mode intensity of split modes also exhibits a different dependence on the gap size in the considered RBs. TMM provides the imaginary part of the Fresnel reflection coefficient, Im{$r$}, that is proportional to the intensity of PhP mode in the heterostructure. The bottom panel in Figure 1D plots the ratio Im{$r_{M1}$}/Im{$r_{M0}$} between the split modes as a function of the gap size for RB-I and RB-II. This ratio is much closer to unity in RB-I, indicating that the mode intensity of the split eigenmodes is more uniform.

Based on the discussion above, we choose RB-I as the target frequency range, since it allows more efficient dispersion manipulation and easier experimental observation of the mode splitting. We start by experimentally verifying the PhP dispersion in the single α-MoO$_3$ crystal and the two adjacent α-MoO$_3$ slabs. To this end, we measure the near-field interference fringes formed by propagating PhPs and then perform a Fourier transform of near-field images to obtain their dispersion.

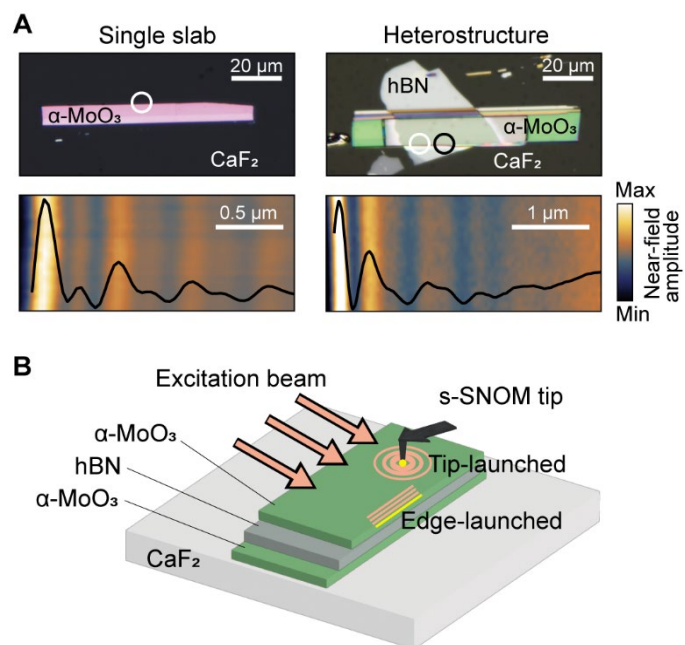

**Figure 2.** **(A)** Optical microscope images of the single-slab sample of thickness 240 nm (top left) and the trilayer heterostructure α-MoO$_3$/hBN/α-MoO$_3$ (249 nm/41 nm/221 nm) on a CaF$_2$ substrate (top right). The white circles indicate the areas of near-field imaging discussed in the main text. The black circle indicates an additional measurement area of the trilayer sample (provided in Supplementary Figure S3A). Bottom: near-field images of the corresponding areas indicated by the white circles; both at an excitation frequency of 985 cm$^{-1}$. **(B)** Schematics of the probed heterostructure where both the sample edge and the SNOM tip launch phonon-polaritons.

The ideal structure would be two suspended α-MoO$_3$ slabs separated by a small air gap. However, we simplify the sample fabrication by replacing the air gap with a thin hBN slab which plays the role of a dielectric spacer. At the same time, we use a CaF$_2$ substrate since its permittivity is close to that of air ($n\sim1.2$ at the frequencies of interest) to preserve the symmetry of the system as much as possible. The top row in Figure 2A shows optical microscope images of the single 240 nm-thick α-MoO$_3$ flake (left) and the trilayer α-MoO$_3$/hBN/α-MoO$_3$ (249/41/221 nm) heterostructure (right) on a CaF$_2$ substrate, along with the near-field images of the edges indicated by white circles (bottom row); a black circle marks the area where additional near-field data is collected, which is discussed in Supplementary Figure S3A. In addition, the results obtained from the second trilayer α-MoO$_3$/hBN/α-MoO$_3$ (238/31/224 nm) sample are provided in Supplementary Figure S4. α-MoO$_3$ and hBN flakes are obtained by exfoliation, and their thicknesses are measured with an atomic force microscope (AFM). All samples are fabricated using the dry transfer method (Supplementary Figure S5A).

Since the momentum of PhPs is much larger than that of free-space light, they cannot be directly excited by the illumination beam. In the s-SNOM experiments, sharp crystal edges and the s-SNOM nano-tip both scatter free-space photons and launch high-momentum PhPs (Figure 2B), typically leading to a mixed near-field signal[18]. Near-field images at a representative frequency of 985 cm$^{-1}$ (the bottom row in Figure 2A) show clear fringes in both samples. In the single-flake sample, periodic fringes are clearly observed, while in the trilayer sample, the fringes appear to be formed by a mix of multiple modes. To quantitatively analyze which modes contribute to the near-field patterns, we analyze the near-field profiles across the frequency range from 975 cm$^{-1}$ to 990 cm$^{-1}$ with a step of 2.5 cm$^{-1}$ (Figure 3A; near-field images are shown in Supplementary Figure S3). Fourier spectra of the profiles (Figure 3B) readily provide the momenta of the corresponding modes that are revealed by spectral peaks.

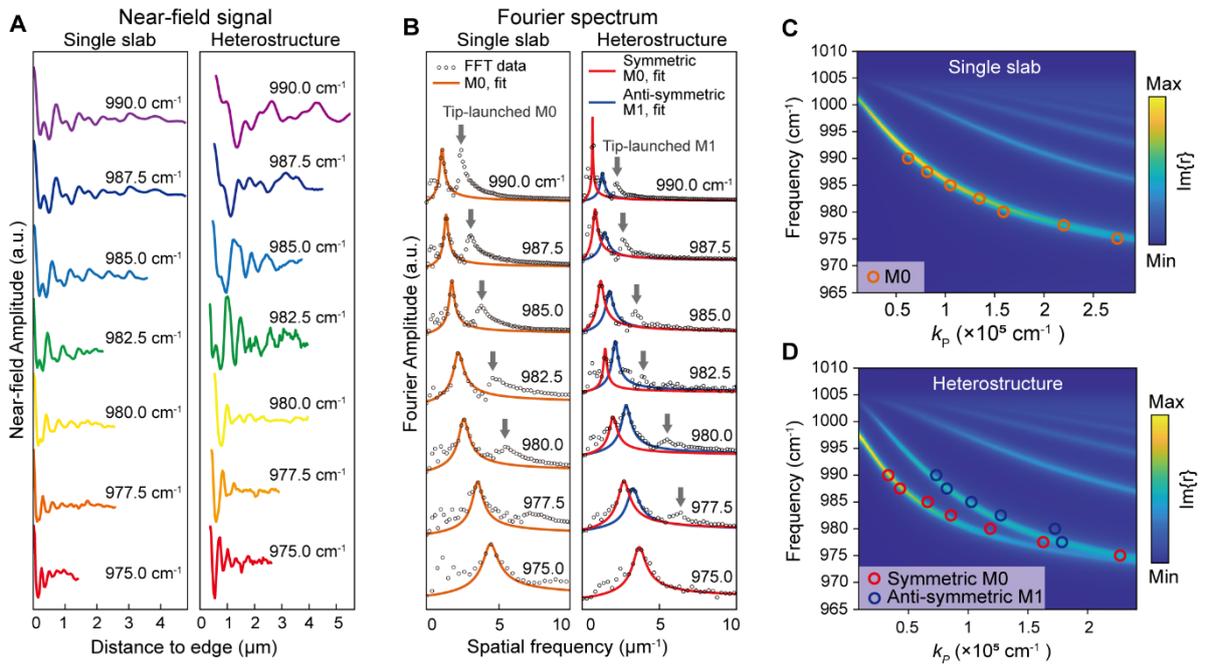

**Figure 3. (A)** Near-field amplitude profiles measured across the edge of the single α-MoO$_3$ slab (left) and the α-MoO$_3$/hBN/α-MoO$_3$ heterostructure (right) at multiple frequencies. **(B)** Spatial Fourier spectra of the corresponding near-field signals shown in (A). Spectra of all edge-launched modes are fitted with a Lorentzian function; gray arrows indicate spectral signal from the tip-launched modes. **(C)** PhP dispersion calculated using the TMM (color plot) and extracted from s-SNOM measurements (orange circles) for the single α-MoO$_3$ slab; experimentally observed mode corresponds to the fundamental mode M0. **(D)** Same as in (C), but for the trilayer heterostructure. Here, the experimentally observed dispersion corresponds to the symmetric mode M0 (red dots) and the anti-symmetric mode M1 (blue dots) resulting from the mode splitting.

For the single flake, two distinct peaks are observed. The momenta of the two peaks maintain an approximately constant ratio of two, corresponding to the edge- and tip-launched mode M0 (left panel in Figure 3B, fitted by the orange Lorentzian). The lower-frequency peak corresponds to the PhP mode excited by the sample edge, while the higher-frequency peak is attributed to the same mode launched by the tip and consequently reflected from the crystal edge, which has been extensively studied before[18-20].

For the trilayer sample, three peaks are constantly present (right panel in Figure 3B), where the two closely located peaks with the highest amplitudes can be attributed to the split symmetric mode (lower-momentum; fitted by the red Lorentzian) and anti-symmetric mode (higher-momentum; fitted by the blue Lorentzian). The relatively weaker third peak with the highest frequency is attributed to the tip-launched anti-symmetric mode, since its momentum linearly scales with that of the anti-symmetric mode with a factor of 2 at different frequencies[18]. We speculate that the tip-sample near-field interaction is favorable for the excitation of the anti-symmetric mode, thus we do not detect the signal from the tip-launched symmetric mode. Figures 3C and D demonstrate the extracted polariton momenta overlapped with the calculated dispersion, showing a very good agreement for all cases.

Additionally, we numerically study several more cases in which dispersion splitting in an ideal α-MoO$_3$/air/α-MoO$_3$ system can be further manipulated. Supplementary Figure S6A shows the dispersion evolution as the air gap becomes thinner, leading to the decreased momentum of the symmetric mode, while the momentum of the anti-symmetric mode increases. As a result, the splitting becomes larger. At the same time, if the thickness of both α-MoO$_3$ slabs ($t_1 = t_2$) increases, the momenta of both symmetric and anti-symmetric modes decrease (Supplementary Figure S6B). This behavior is expected since the split modes originate from the eigenmodes of a single slab which are inversely proportional to the slab thickness[21].

Furthermore, by tuning the thicknesses in the system, it is possible to selectively increase or decrease the momentum of only one of the split modes, while keeping the momentum of the other mode almost unchanged. For example, Supplementary Figure S6C shows the dispersion evolution when the symmetry between the top and bottom α-MoO$_3$ layers is broken while both gap and $t_1 + t_2$ are constant. We note that when the structure is asymmetric, the field topology of the modes changes compared to the symmetric case (Supplementary Figure S7). Finally, when the gap thickness is varied in the asymmetric structure with $t_2 > t_1$, the lower-momentum mode converges to the dispersion of the thicker slab, while the higher-momentum mode converges to that of the thinner slab (Supplementary Figure S6D, bottom panel). Since the TMM assumes the reflection coefficient measured from the top half-space, the intensity of the mode corresponding to the bottom slab becomes weaker due to the reduced evanescent coupling as the gap increases.

In the final part of this study, we employ graphene to achieve an electrically-driven dispersion manipulation in the heterostructure. The in-plane conductivity of graphene strongly depends on the Fermi level[22], which can be dynamically tuned via electrostatic gating[23-26]. By leveraging this tunability and the field symmetry of the modes, a selective and dynamic dispersion control of the polaritons can be achieved.

To demonstrate this, we consider a symmetric heterostructure on CaF$_2$ substrate schematically shown in Figure 4A, where an hBN-encapsulated graphene is sandwiched between two identical α-MoO$_3$ slabs. The two lowest-order polariton modes in this structure are the symmetric M0 mode and the anti-symmetric M1 mode. In contrast to the out-of-plane field, $E_z$, the distribution of the in-plane field, $E_x$, of the M0 mode is antisymmetric with respect to the graphene plane and practically vanishes at the position of graphene (Figure 4B), whereas the M1 mode possesses symmetric $E_x$ profile with the maximum occurring very close to graphene plane (Figure 4C). As a consequence, the former does not *feel* the graphene but the latter is susceptible to the graphene conductivity modulation.

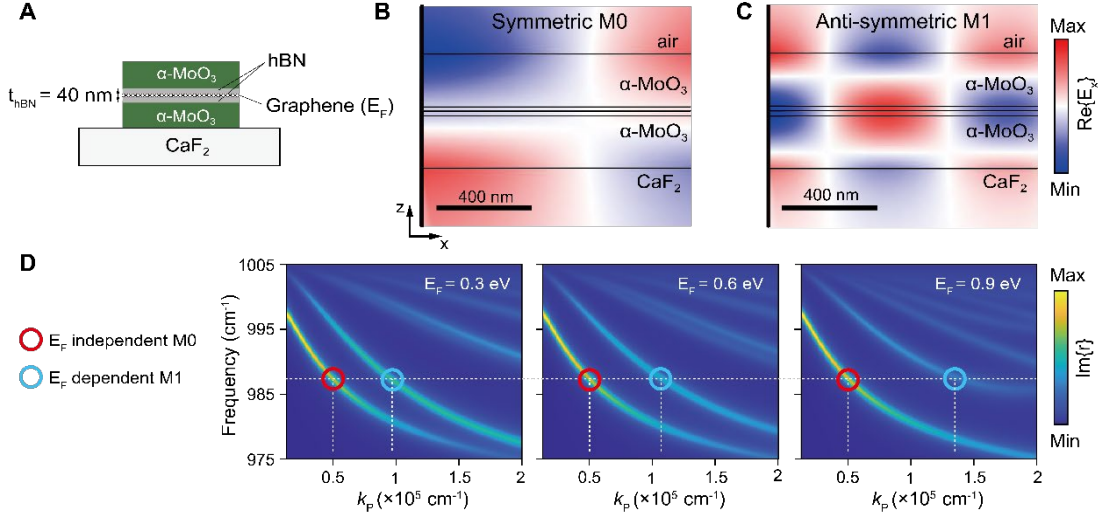

**Figure 4. (A)** Heterostructure with the hBN-encapsulated graphene placed between the α-MoO$_3$ slabs. **(B)** Spatial distribution of the in-plane electric field of the M0 mode in the heterostructure shown in (A) with the graphene's Fermi energy of 0.6 eV; at 995 cm$^{-1}$. **(C)** Same as in (B), but for the M1 mode in the heterostructure. **(D)** PhP dispersion calculated for the five-layer α-MoO$_3$/hBN/graphene/hBN/α-MoO$_3$/CaF$_2$ structure with the Fermi energy of graphene assumed as 0.3 eV (left), 0.6 eV (middle), and 0.9 eV (right). Red (blue) circles indicate the mode that is insensitive (sensitive) to the Fermi energy; white dashed lines are a guide to the eye. All α-MoO$_3$ slabs have a thickness of 230 nm and the thickness of each hBN layer is 20 nm.

The Fermi level-dependent PhP dispersion is shown in Figure 4D. As the Fermi level of graphene increases from 0.3 to 0.9 eV, the dispersion of the M0 mode remains unaffected, while the M1 mode blueshifts. In terms of the mode confinement at a fixed frequency of 995 cm$^{-1}$, the momentum of the M1 mode increases by 1.3 times for the assumed Fermi level modulation. At the same time, the increased graphene conductivity makes the M1 mode more dissipative. We also investigate the role of the hBN spacer which encapsulates graphene (Supplementary Figures S8A and B), showing that the dispersion of split modes can be systematically engineered by tuning the graphene Fermi level and the hBN thickness, as illustrated in Supplementary Figure S8C.

**Conclusion**

We experimentally demonstrate the phonon-polariton eigenmode splitting into symmetric and antisymmetric branches by near-field imaging of the α-MoO$_3$/hBN/α-MoO$_3$ heterostructure using s-SNOM. By quantitatively comparing the polariton dispersion predicted by the TMM with the experimental results, the mode splitting has been verified for the polaritons in the Type-I Reststrahlen band of α-MoO$_3$. Furthermore, we numerically show how tuning the gap (spacer) size between the two α-MoO$_3$ slabs or their thickness provides a powerful method to achieve dispersion engineering for the symmetric and antisymmetric modes, which can be applied to other polaritonic materials. In addition, we numerically demonstrate the tunable and mode-selective dispersion engineering by inserting graphene into the gap between the two α-MoO$_3$ crystals and assuming different Fermi levels.

AUTHOR CONTRIBUTION

D. N. and J. O. fabricated the samples, performed near-field measurements, and analyzed the data. D. N. wrote the manuscript and performed numerical simulations. S. G. M. and M. S. J. proposed the research idea, contributed to manuscript writing, and supervised the project.

ACKNOWLEDGMENTS

This research was supported by the National Research Foundation of Korea (NRF) grants funded by the Ministry of Science, ICT and Future Planning (NRF-RS-2022-NR070476 and NRF-RS-2024-00340639). This work was also supported by the BK21 FOUR Program through the NRF funded by Ministry of Education.

REFERENCES

1. Galiffi, E.; Carini, G.; Ni, X.; Álvarez-Pérez, G.; Yves, S.; Renzi, E. M.; Nolen, R.; Wasserroth, S.; Wolf, M.; Alonso-Gonzalez, P.; Paarmann, A.; Alù, A., Extreme light confinement and control in low-symmetry phonon-polaritonic crystals. *Nat. Rev. Mater.* **2023,** *9* (1), 9-28.
2. Ma, W.; Alonso-Gonzalez, P.; Li, S.; Nikitin, A. Y.; Yuan, J.; Martin-Sanchez, J.; Taboada-Gutierrez, J.; Amenabar, I.; Li, P.; Velez, S.; Tollan, C.; Dai, Z.; Zhang, Y.; Sriram, S.; Kalantar-Zadeh, K.; Lee, S. T.; Hillenbrand, R.; Bao, Q., In-plane anisotropic and ultra-low-loss polaritons in a natural van der Waals crystal. *Nature* **2018,** *562* (7728), 557-562.
3. Caldwell, J. D.; Kretinin, A. V.; Chen, Y.; Giannini, V.; Fogler, M. M.; Francescato, Y.; Ellis, C. T.; Tischler, J. G.; Woods, C. R.; Giles, A. J.; Hong, M.; Watanabe, K.; Taniguchi, T.; Maier, S. A.; Novoselov, K. S., Sub-diffractional volume-confined polaritons in the natural hyperbolic material hexagonal boron nitride. *Nat. Commun.* **2014,** *5*, 5221.
4. Álvarez-Pérez, G.; Folland, T. G.; Errea, I.; Taboada-Gutierrez, J.; Duan, J.; Martin-Sanchez, J.; Tresguerres-Mata, A. I. F.; Matson, J. R.; Bylinkin, A.; He, M.; Ma, W.; Bao, Q.; Martin, J. I.; Caldwell, J. D.; Nikitin, A. Y.; Alonso-Gonzalez, P., Infrared Permittivity of the Biaxial van der Waals Semiconductor α-MoO$_3$ from Near- and Far-Field Correlative Studies. *Adv. Mater.* **2020,** *32* (29), 1908176.

5. Passler, N. C.; Ni, X.; Hu, G.; Matson, J. R.; Carini, G.; Wolf, M.; Schubert, M.; Alu, A.; Caldwell, J. D.; Folland, T. G.; Paarmann, A., Hyperbolic shear polaritons in low-symmetry crystals. *Nature* **2022**, *602* (7898), 595-600.
6. Hu, G.; Ou, Q.; Si, G.; Wu, Y.; Wu, J.; Dai, Z.; Krasnok, A.; Mazor, Y.; Zhang, Q.; Bao, Q.; Qiu, C. W.; Alu, A., Topological polaritons and photonic magic angles in twisted α-MoO₃ bilayers. *Nature* **2020**, *582* (7811), 209-213.
7. Menabde, S. G.; Boroviks, S.; Ahn, J.; Heiden, J. T.; Watanabe, K.; Taniguchi, T.; Low, T.; Hwang, D. K.; Mortensen, N. A.; Jang, M. S., Near-field probing of image phonon-polaritons in hexagonal boron nitride on gold crystals. *Sci. Adv.* **2022**, *8* (28), eabn0627.
8. Menabde, S. G.; Jahng, J.; Boroviks, S.; Ahn, J.; Heiden, J. T.; Hwang, D. K.; Lee, E. S.; Mortensen, N. A.; Jang, M. S., Low-Loss Anisotropic Image Polaritons in van der Waals Crystal α-MoO₃. *Adv. Opt. Mater.* **2022**, *10* (21), 2201492.
9. Lee, I. H.; He, M.; Zhang, X.; Luo, Y.; Liu, S.; Edgar, J. H.; Wang, K.; Avouris, P.; Low, T.; Caldwell, J. D.; Oh, S. H., Image polaritons in boron nitride for extreme polariton confinement with low losses. *Nat. Commun.* **2020**, *11* (1), 3649.
10. Menabde, S. G.; Lim, Y.; Voronin, K.; Heiden, J. T.; Nikitin, A. Y.; Lee, S.; Jang, M. S., Polaritonic Fourier crystal. *Nat. Commun.* **2025**, *16* (1), 2530.
11. Menabde, S. G.; Lim, Y.; Nikitin, A. Y.; Alonso-González, P.; Heiden, J. T.; Noh, H.; Lee, S.; Jang, M. S., Quasi-Single-Mode Polaritonic Crystal for Hyperbolic Phonon-Polaritons. *Adv. Opt. Mater.* **2026**, e02662.
12. Hu, H.; Chen, N.; Teng, H.; Yu, R.; Qu, Y.; Sun, J.; Xue, M.; Hu, D.; Wu, B.; Li, C.; Chen, J.; Liu, M.; Sun, Z.; Liu, Y.; Li, P.; Fan, S.; Garcia de Abajo, F. J.; Dai, Q., Doping-driven topological polaritons in graphene/α-MoO₃ heterostructures. *Nat. Nanotechnol.* **2022**, *17* (9), 940-946.
13. Tu, P. Y.; Huang, C. C., Analysis of hybrid plasmon-phonon-polariton modes in hBN/graphene/hBN stacks for mid-infrared waveguiding. *Opt. Express* **2022**, *30* (2), 2863-2876.
14. Zheng, Y.; Zeng, Y.; Hu, Y.; Liu, L.; Dai, Z.; Chen, H., Hyperbolic polaritons in twisted β-Ga₂O₃. *npj Nanophoton.* **2025**, *2* (1), 10.
15. Wu, Y.; Ou, Q.; Yin, Y.; Li, Y.; Ma, W.; Yu, W.; Liu, G.; Cui, X.; Bao, X.; Duan, J.; Alvarez-Perez, G.; Dai, Z.; Shabbir, B.; Medhekar, N.; Li, X.; Li, C. M.; Alonso-Gonzalez, P.; Bao, Q., Chemical switching of low-loss phonon polaritons in α-MoO₃ by hydrogen intercalation. *Nat. Commun.* **2020**, *11* (1), 2646.
16. Passler, N. C.; Paarmann, A., Generalized 4 × 4 matrix formalism for light propagation in anisotropic stratified media: study of surface phonon polaritons in polar dielectric heterostructures: erratum. *J. Opt. Soc. Am. B* **2019**, *36* (11), 3246–3248.
17. Yariv, A.; Yeh, P. *Optical Electronics in Modern Communications,* 6th edition; Oxford University Press: 2007.
18. Jang, M.; Menabde, S. G.; Kiani, F.; Heiden, J. T.; Zenin, V. A.; Asger Mortensen, N.; Tagliabue, G.; Jang, M. S., Fourier analysis of near-field patterns generated by propagating polaritons. *Phys. Rev. Appl.* **2024**, *22* (1), 014076.
19. Huber, A.; Ocelic, N.; Kazantsev, D.; Hillenbrand, R., Near-field imaging of mid-

infrared surface phonon polariton propagation. *Appl. Phys. Lett.* **2005,** *87* (8), 081103.
20. Dai, S.; Ma, Q.; Yang, Y.; Rosenfeld, J.; Goldflam, M. D.; McLeod, A.; Sun, Z.; Andersen, T. I.; Fei, Z.; Liu, M.; Shao, Y.; Watanabe, K.; Taniguchi, T.; Thiemens, M.; Keilmann, F.; Jarillo-Herrero, P.; Fogler, M. M.; Basov, D. N., Efficiency of Launching Highly Confined Polaritons by Infrared Light Incident on a Hyperbolic Material. *Nano Lett.* **2017,** *17* (9), 5285-5290.
21. Álvarez-Pérez, G.; Voronin, K. V.; Volkov, V. S.; Alonso-González, P.; Nikitin, A. Y., Analytical approximations for the dispersion of electromagnetic modes in slabs of biaxial crystals. *Phys. Rev. B* **2019,** *100* (23), 235408.
22. Koppens, F. H.; Chang, D. E.; Garcia de Abajo, F. J., Graphene plasmonics: a platform for strong light-matter interactions. *Nano Lett.* **2011,** *11* (8), 3370-7.
23. Brar, V. W.; Jang, M. S.; Sherrott, M.; Lopez, J. J.; Atwater, H. A., Highly confined tunable mid-infrared plasmonics in graphene nanoresonators. *Nano Lett.* **2013**, *13*, 2541–2547.
24. Kim, S.; Jang, M. S.; Brar, V. W.; Mauser, K. W.; Kim, L.; Atwater, H. A., Electronically tunable perfect absorption in graphene. *Nano Lett.* **2018**, *18*, 971–979.
25. Siegel, J.; Kim, S.; Fortman, M.; Wan, C.; Kats, M. A.; Hon, P. W. C.; Sweatlock, L.; Jang, M. S.; Brar, V. W., Electrostatic steering of thermal emission with active metasurface control of delocalized modes. *Nat. Commun.* **2024**, *15*, 3376.
26. Han, S.; Kong, J.; Choi, J.; Chegal, W.; Jang, M. S., Single-gate electro-optic beam switching metasurfaces. *Light Sci. Appl.* **2025**, *14*, 292.